\documentclass{elsart}
\usepackage{graphicx}
\usepackage{amssymb}
\begin{document}

\begin{frontmatter}

\title{A Search for Arrival Direction Clustering in the HiRes-I Monocular 
Data above {\boldmath$10^{19.5}$}~eV} 

\author[utah]{R.U.~Abbasi,}
\author[utah]{T.~Abu-Zayyad,}
\author[lanl]{J.F.~Amann,}
\author[utah]{G.~Archbold,}
\author[utah]{R.~Atkins,}
\author[adelaide]{J.A.~Bellido,}
\author[utah]{K.~Belov,}
\author[umt]{J.W.~Belz,}
\author[nevis]{S.~BenZvi,}
\author[rutgers]{D.R.~Bergman,}
\author[utah]{G.W.~Burt,}
\author[utah]{Z.~Cao,}
\author[adelaide]{R.W.~Clay,}
\author[nevis]{B.~Connolly,}
\author[adelaide]{B.R.~Dawson,}
\author[utah]{W.~Deng,}
\author[utah]{Y.~Fedorova,}
\author[utah]{J.~Findlay,}
\author[nevis]{C.B.~Finley,}
\author[utah]{W.F.~Hanlon,}
\author[lanl]{C.M.~Hoffman,}
\author[lanl]{M.H.~Holzscheiter,}
\author[rutgers]{G.A.~Hughes,}
\author[utah]{P.~H\"{u}ntemeyer,}
\author[utah]{C.C.H.~Jui,}
\author[utah]{K.~Kim,}
\author[umt]{M.A.~Kirn,}
\author[utah]{E.C.~Loh,}
\author[utah]{M.M.~Maestas,}
\author[icrr]{N.~Manago,}
\author[lanl]{L.J.~Marek,}
\author[utah]{K.~Martens,}
\author[unm]{J.A.J.~Matthews,}
\author[utah]{J.N.~Matthews,}
\author[nevis]{A.~O'Neill,}
\author[lanl]{C.A.~Painter,}
\author[rutgers]{L.~Perera,}
\author[utah]{K.~Reil,}
\author[utah]{R.~Riehle,}
\author[unm]{M.~Roberts,}
\author[icrr]{M.~Sasaki,}
\author[rutgers]{S.R.~Schnetzer,}
\author[adelaide]{K.M.~Simpson,}
\author[lanl]{G.~Sinnis,}
\author[utah]{J.D.~Smith,}
\author[utah]{R.~Snow,}
\author[utah]{P.~Sokolsky,}
\author[nevis]{C.~Song,}
\author[utah]{R.W.~Springer,}
\author[utah]{B.T.~Stokes\corauthref{cor1},}
\author[utah]{J.R.~Thomas,}
\author[utah]{S.B.~Thomas,}
\author[rutgers]{G.B.~Thomson,}
\author[lanl]{D.~Tupa,}
\author[nevis]{S.~Westerhoff,}
\author[utah]{L.R.~Wiencke,}
\author[rutgers]{and A.~Zech}
\collaboration{The High Resolution Fly's Eye Collaboration}
\address[utah]{University of Utah,
Department of Physics and High Energy Astrophysics Institute,
Salt Lake City, UT~84112, USA}
\address[lanl]{Los Alamos National Laboratory,
Los Alamos, NM~87545, USA}
\address[adelaide]{University of Adelaide, Department of Physics,
Adelaide, SA~5005, Australia}
\address[umt]{University of Montana, Department of Physics and Astronomy,
Missoula, MT~59812, USA.}
\address[nevis]{Columbia University, Department of Physics and
Nevis Laboratories, New York, NY~10027, USA}
\address[rutgers]{Rutgers --- The State University of New Jersey,
Department of Physics and Astronomy,
Piscataway, NJ~08854, USA}
\address[icrr]{University of Tokyo,
Institute for Cosmic Ray Research,
Kashiwa City, Chiba~277-8582, Japan}
\address[unm]{University of New Mexico,
Department of Physics and Astronomy,
Albuquerque, NM~87131, USA}
\corauth[cor1]{
Corresponding~author.\ {\it E-mail~address}:~stokes@cosmic.utah.edu~(B.T.~Stokes)
}

\newpage

\begin{abstract}
In the past few years, small scale anisotropy has become a primary focus in
the search for source of Ultra-High Energy 
Cosmic Rays (UHECRs).  The Akeno Giant Air
Shower Array (AGASA) has reported the presence of clusters of event arrival
directions in their highest energy data set.  
The High Resolution Fly's Eye (HiRes)
has accumulated an exposure in one of its monocular eyes at energies 
above $10^{19.5}$~eV comparable 
to that of AGASA.  However, monocular events observed with an air fluorescence
detector are characterized by highly asymmetric angular resolution.  A method
is developed for measuring autocorrelation with asymmetric angular resolution.
It is concluded that HiRes-I observations are consistent with no 
autocorrelation and that the sensitivity to clustering of the HiRes-I detector
is comparable to that of the reported AGASA data set.  Furthermore, we state
with a 90\% confidence level that no more than 13\% of the observed HiRes-I
events above $10^{19.5}$~eV could be sharing common arrival directions. 
However, because a
measure of autocorrelation makes no assumption of the underlying astrophysical
mechanism that results in clustering phenomena, we cannot claim that the HiRes 
monocular analysis and the AGASA analysis are inconsistent beyond a specified
confidence level. 
\end{abstract}

\end{frontmatter}

\section{Introduction}

Over the past decade, the search for sources of Ultra-High Energy 
Cosmic Rays (UHECRs) has begun to focus upon small scale anisotropy in event
arrival directions.  
This refers to statistically significant excesses occurring at the
scale of $\leq2.5^\circ$.  The interest in this sort of anisotropy has
largely been fueled by the observations of the Akeno Giant Air Shower Array 
(AGASA).  In 1999 \cite{Takeda:1999sg} and again in 
2001 \cite{Takeda:2001}, the AGASA collaboration reported observing
what eventually became seven clusters (six ``doublets'' and one ``triplet'') 
with estimated energies above $\sim3.8\times10^{19}$~eV.  
Several attempts that have been made to 
ascertain the significance of these clusters returned chance probabilities
of $4\times10^{-6}$ \cite{Tinyakov:2001ic} to 0.08 
\cite{Finley:2003on}.

By contrast, the monocular (and stereo) analyses that have been 
presented by the High Resolution Fly's
Eye (HiRes) demonstrate that the level of autocorrelation observed 
in our sample is completely consistent with that expected from background
coincidences \cite{bellido,bellido2,apj}.  
Any analysis of HiRes monocular data needs to take into account that the 
angular resolution in monocular mode is highly asymmetric.  

It is very difficult to compare the results of the HiRes
monocular and AGASA analyses.  They are 
very different in the way that they measure autocorrelation.  
Differences in the published energy spectra of the two experiments suggest
an energy scale difference of 30\% \cite{prl,Takeda:1998ps}.
Additionally, the two experiments observe UHECRs in very 
different ways.  The HiRes experiment has an energy-dependent aperture and
an exposure with a seasonal variability \cite{prl}.  These differences
make it very difficult get an intuitive grasp of what HiRes should see if
the AGASA claim of autocorrelation is justified.   In order to develop this
sort of intuition, we apply the same analysis to both AGASA and HiRes data.  

\section{The HiRes-I Monocular Data}

The data set that we consider consists of events
that were included in the HiRes-I monocular spectrum measurement 
\cite{prl,fadc}.
This set contains 52 events observed between May~1997 and 
February~2003 with measured energies greater than $10^{19.5}$~eV.  
The data set represents a cumulative 
exposure of $\sim3000$~km$^2\cdot$sr$\cdot$yr at $5\times10^{19}$~eV.
This data was subject to a number of quality cuts that are detailed in the
above-mentioned papers \cite{prl,fadc}.  We previously verified that this data
set was consistent with Monte Carlo predictions in many ways including
 impact parameter 
($R_p$) distributions \cite{prl} and zenith angle distributions \cite{dipole}. 
For this study, we presumed an average atmospheric clarity
\cite{Wiencke:atmos}.

\begin{figure}[t,b]
\begin{center}
\begin{tabular}{c}
\includegraphics[width=7.0cm]{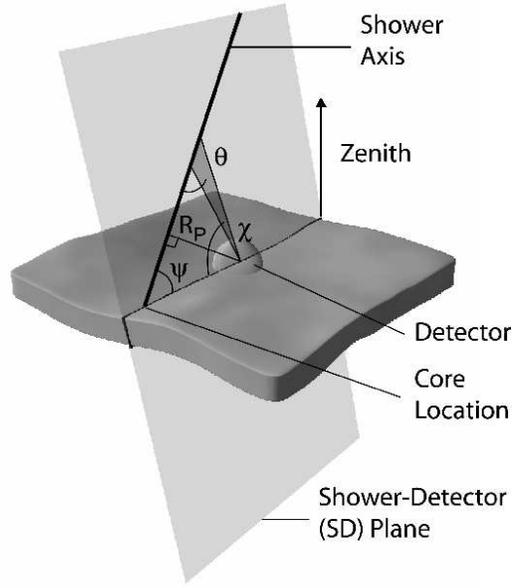}\\
\end{tabular}
\end{center}
\caption{The geometry of reconstruction for a monocular air fluorescence 
detector}
\label{figure:picture}
\end{figure}

In order to calculate the autocorrelation function for this subset of data,
we must first parameterize the HiRes-I monocular angular resolution.
For a monocular air fluorescence detector, angular resolution consists
of two components, the plane of reconstruction, that is the plane in which
the shower is observed, and the angle $\psi$ within
the plane of reconstruction (see figure~\ref{figure:picture}). 
We can determine the plane of reconstruction 
very accurately.  However, the value 
of $\psi$ is more difficult to determine accurately because it is
dependent on the precise results of the profile-constrained fit 
\cite{prl,fadc}.

\begin{figure}[t,b]
\begin{center}
\begin{tabular}{c}
\includegraphics[width=13.5cm]{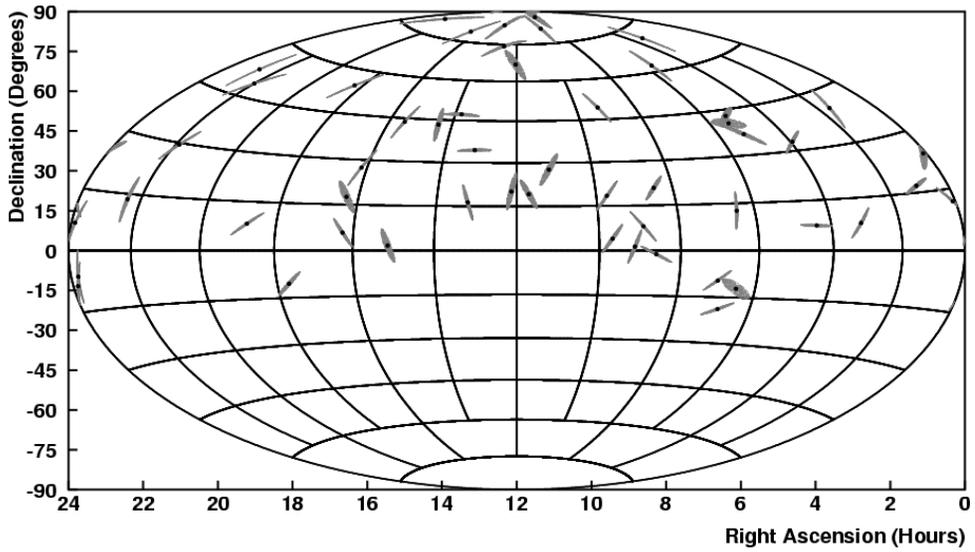}\\
\end{tabular}
\end{center}
\caption{The arrival directions of the HiRes-I monocular with reconstructed 
energies above $10^{19.5}$~eV events and their $1\sigma$ angular resolution}
\label{fig:hires_map}
\end{figure}

The HiRes-I angular resolution is therefore 
described by an elliptical, two-dimensional 
Gaussian distribution with the two Gaussian parameters, $\sigma_\psi$ and 
$\sigma_{\rm plane}$, being defined by the two angular resolutions.
For the range
of estimated energies considered in this paper, 
$\sigma_\psi=[4.9,6.1]^\circ$ and $\sigma_{\rm plane}[0.4,1.5]^\circ$.
In figure~\ref{fig:hires_map}, 
the arrival directions of the
HiRes-I events  are plotted in
equatorial coordinates along with their $1\sigma$ error ellipses.

\begin{figure}[t,b,p]
\begin{center}
\begin{tabular}{c}
\includegraphics[width=8.5cm]{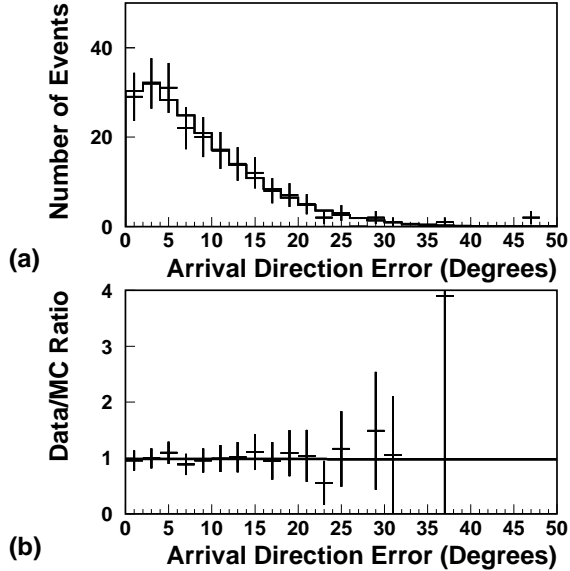}\\
\end{tabular}
\end{center}
\caption{Arrival direction error comparison between real data (mono vs. stereo)
and simulated data for events with estimated energies above $10^{18.5}$~eV.
The solid line histogram corresponds to the 
arrival direction error distribution of the monocular reconstructed 
Monte Carlo simulated data.  
The crosses correspond to the arrival directions error 
distribution observed for actual data by comparing the arrival directions 
estimated by the monocular and stereo reconstructions.  The solid line in the 
ratio component corresponds to the fit $y=ax+b$ where $a=0.000\pm0.011$ and
$b=0.98\pm0.11$.}
\label{fig:comp}
\end{figure}

In order to understand the systematic uncertainty in the angular resolution 
estimates, we 
consider a comparison of estimated arrival directions that successfully
reconstructed in both HiRes-I monocular mode and HiRes stereo mode.  
Because of the dearth of events with estimated energies above $10^{19.5}$~eV
that reconstructed satisfactorily in both stereo and mono mode, we consider 
all mono/stereo candidate events with estimated energies above $10^{18.5}$~eV.
In stereo mode, the shower detector planes of the two detectors are 
intersected, thus the geometry is much more precisely known and the total
angular resolution is of order $0.6^\circ$, a number that is
largely correlated to $\sigma_{\rm plane}$ and thus is negligible
when added in quadrature to the larger term, $\sigma_\psi$.  This allows us
to perform a comparison of the angular resolution estimated through simulations
to the observed angular resolution values of actual data.
In figure~\ref{fig:comp},
we show the distribution of angular errors for real and simulated data.  
The uncertainty in the slope of the ratio (figure~\ref{fig:comp}b) leads to 
an 7.5\% uncertainty in the angular resolution.

\section{The Published AGASA Data}

\begin{figure}[t,b]
\begin{center}
\begin{tabular}{c}
\includegraphics[width=8.0cm]{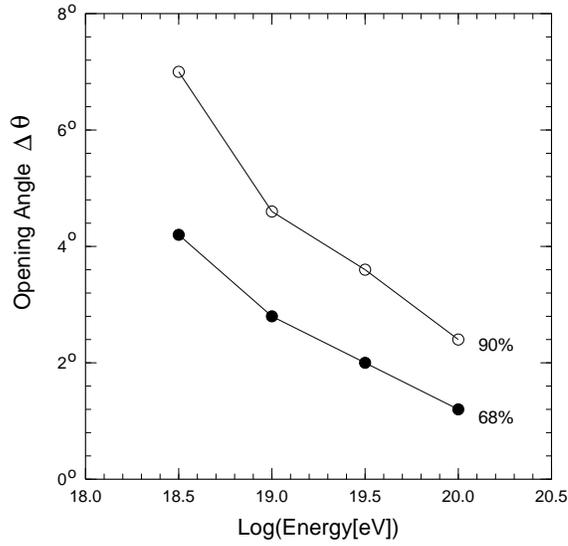}\\
\end{tabular}
\end{center}
\caption{The AGASA angular resolution as a function of estimated energy
\cite{Takeda:1999sg}}
\label{fig:angres}
\end{figure}

\begin{figure}[t,b]
\begin{center}
\begin{tabular}{c}
\includegraphics[width=13.5cm]{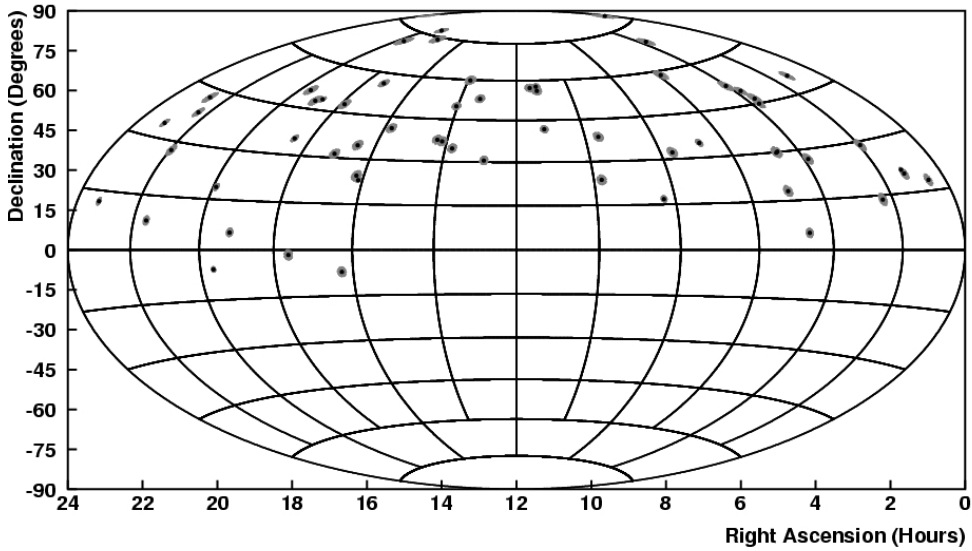}\\
\end{tabular}
\end{center}
\caption{The arrival directions of the published AGASA events with their
68\% angular resolution}
\label{fig:agasa_map}
\end{figure}

The AGASA data with energies above 40~EeV has been published up to the year
2000 \cite{Takeda:2001} and all but one of 
these events used for this calculation has a measured energy greater than
$4\times10^{19}$~eV.
The AGASA estimated angular errors \cite{Takeda:1999sg} are shown in 
figure~\ref{fig:angres}.
The AGASA angular errors (figure~\ref{fig:angres}) are fit to a 
two-component Gaussian distribution:
\begin{equation}
n=N_\circ(E_{\rm EeV})\biggr[0.33\Delta\theta e^{-(\Delta\theta)^2/2\sigma_1^2}
+0.67\Delta\theta e^{-(\Delta\theta)^2/2\sigma_2^2}\biggr]
\label{eqn:angres}
\end{equation}
where $\sigma_1=6.52^\circ-2.16^\circ\log_{10}E_{\rm EeV}$, 
$\sigma_2=3.25^\circ-1.22^\circ\log_{10}E_{\rm EeV}$,
and $N_\circ(E)$ is a numerically determined 
normalization constant. Figure~\ref{fig:agasa_map}
shows the arrival directions of the published
AGASA events plotted in equatorial
coordinates with their 68\% angular resolution.

\section{The Autocorrelation Function}

We measure the degree of autocorrelation in both samples by means
of an autocorrelation function.  It is calculated as follows:
\begin{enumerate}
\item For each event, an arrival direction is sampled on a probabilistic basis
from the error space defined by the  angular resolution of the event.
\item The opening angle is measured between the arrival directions of a pair
of events.
\item The cosine of the opening angle is then histogrammed.
\item The preceding steps are repeated until all possible pairs of the 
events are considered.
\item The preceding steps are repeated until the error space, in the arrival
direction of each event, is thoroughly sampled.
\item The histogram is normalized and the resulting curve is the 
autocorrelation function.
\end{enumerate}

\begin{figure}[t,b]
\begin{center}
\begin{tabular}{c@{\hspace{0.0cm}}c}
(a)\includegraphics[width=6.15cm]{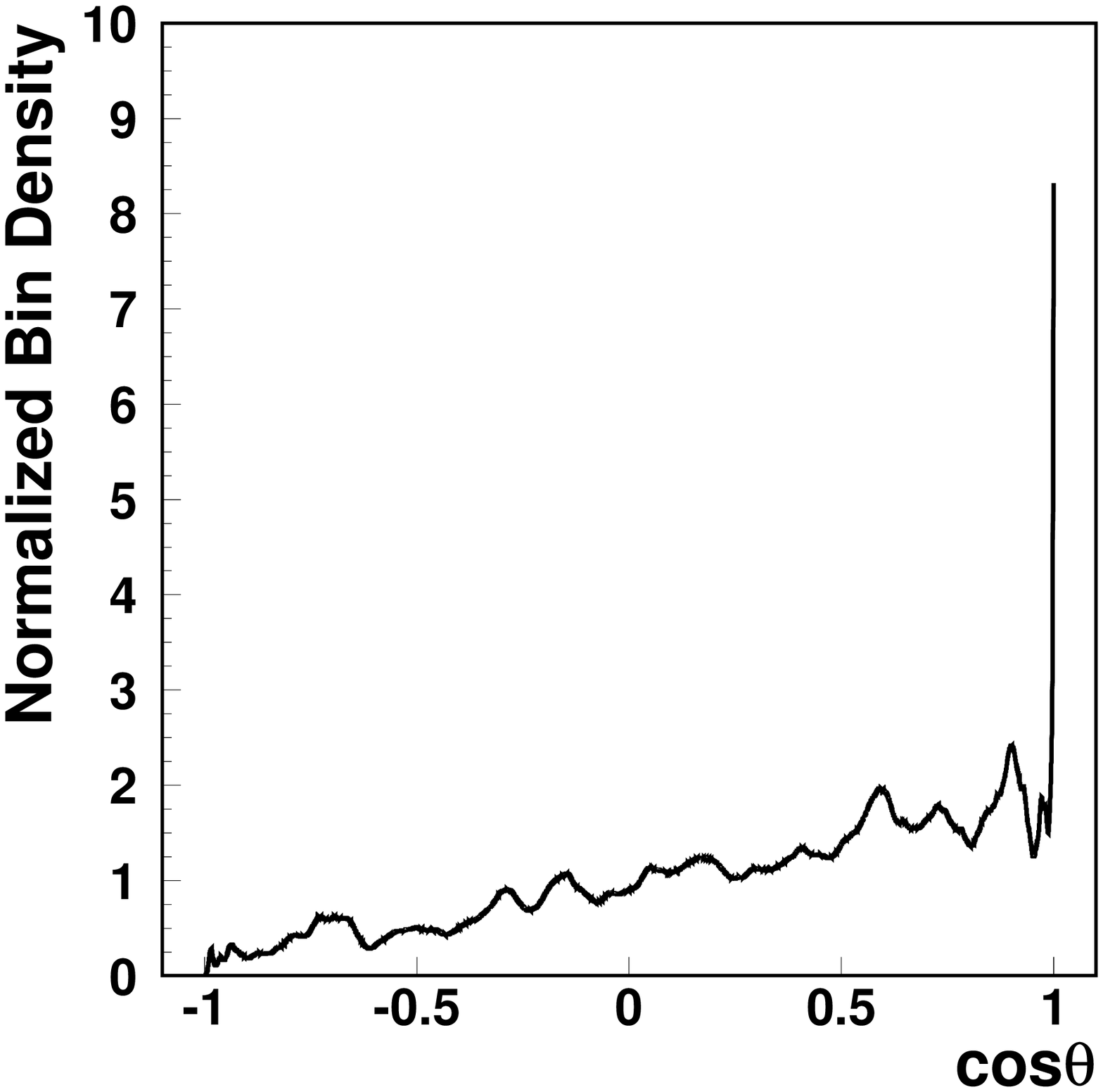}&
(b)\includegraphics[width=6.15cm]{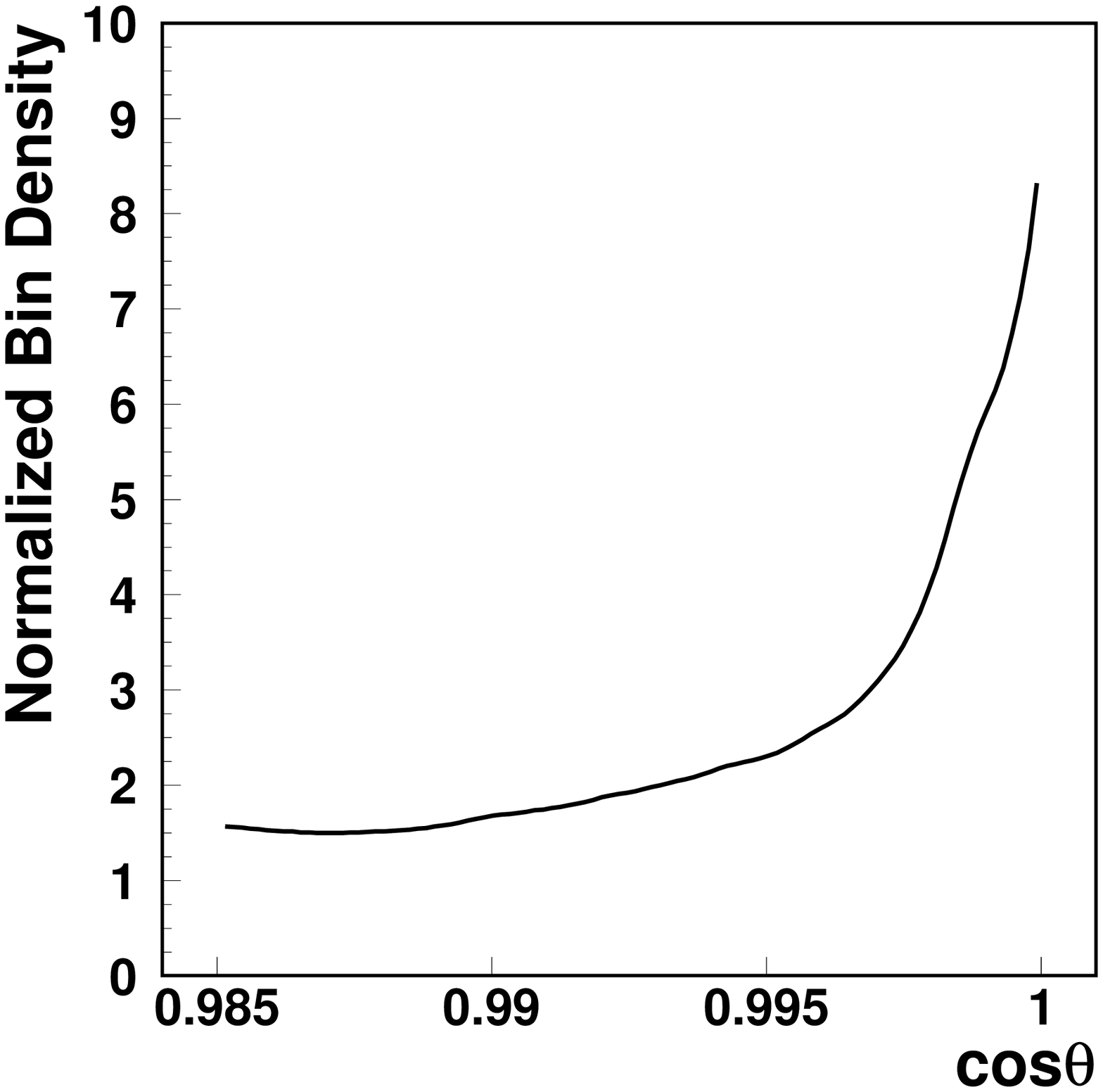}\\
\end{tabular}
\end{center}
\caption{An example of the autocorrelation function for a simulated data set 
that contains $\sim10$ clusters in a total of 60 events---(a) the full 
autocorrelation function for $\theta=[0^\circ,180^\circ]$; (b) the critical
region of the the autocorrelation function: $\theta=[0^\circ,10^\circ]$.}
\label{fig:autoex}
\end{figure}

\begin{figure}[t,b]
\begin{center}
\begin{tabular}{c@{\hspace{0.0cm}}c}
(a)\includegraphics[width=6.15cm]{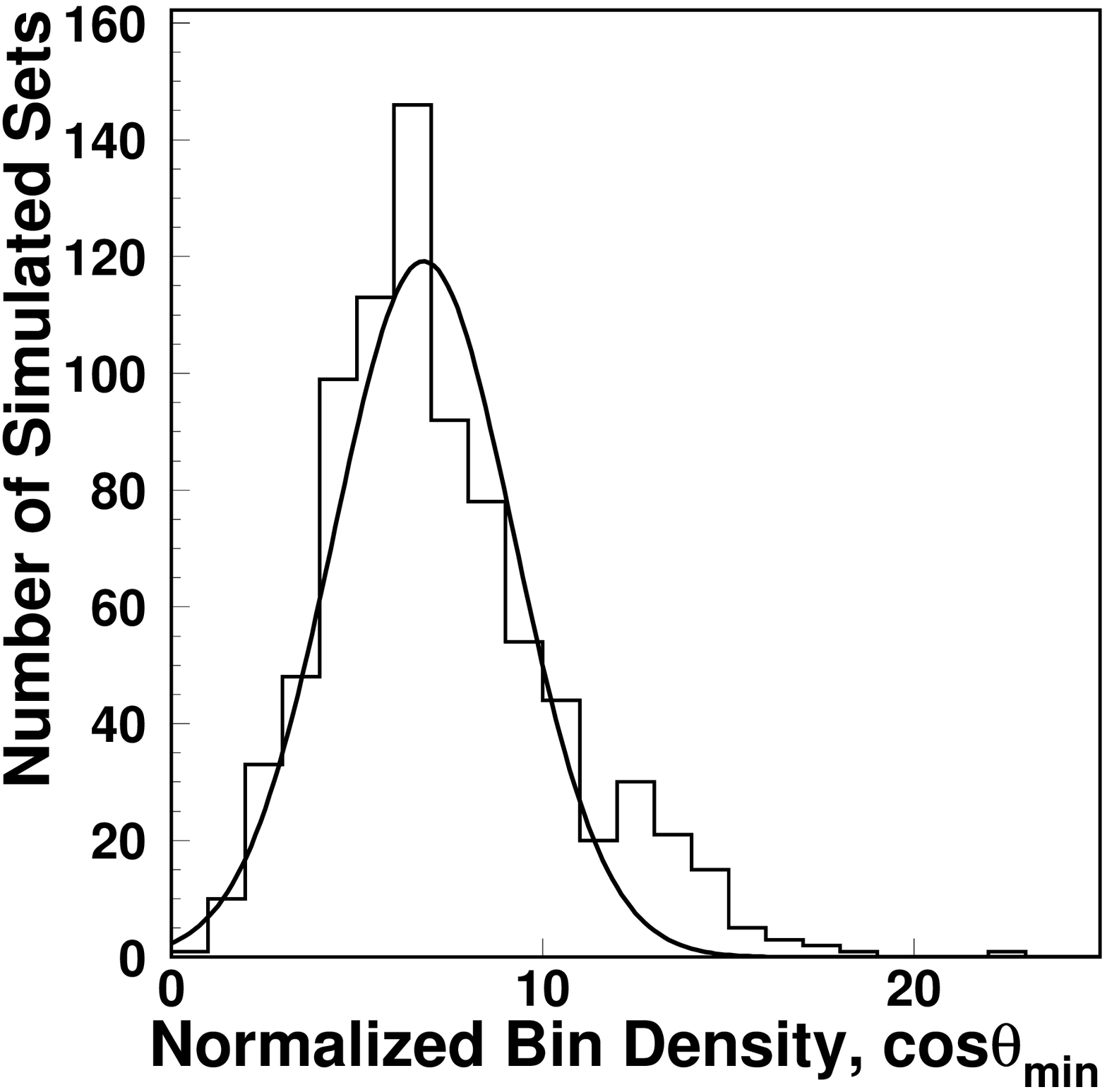}&
(b)\includegraphics[width=6.15cm]{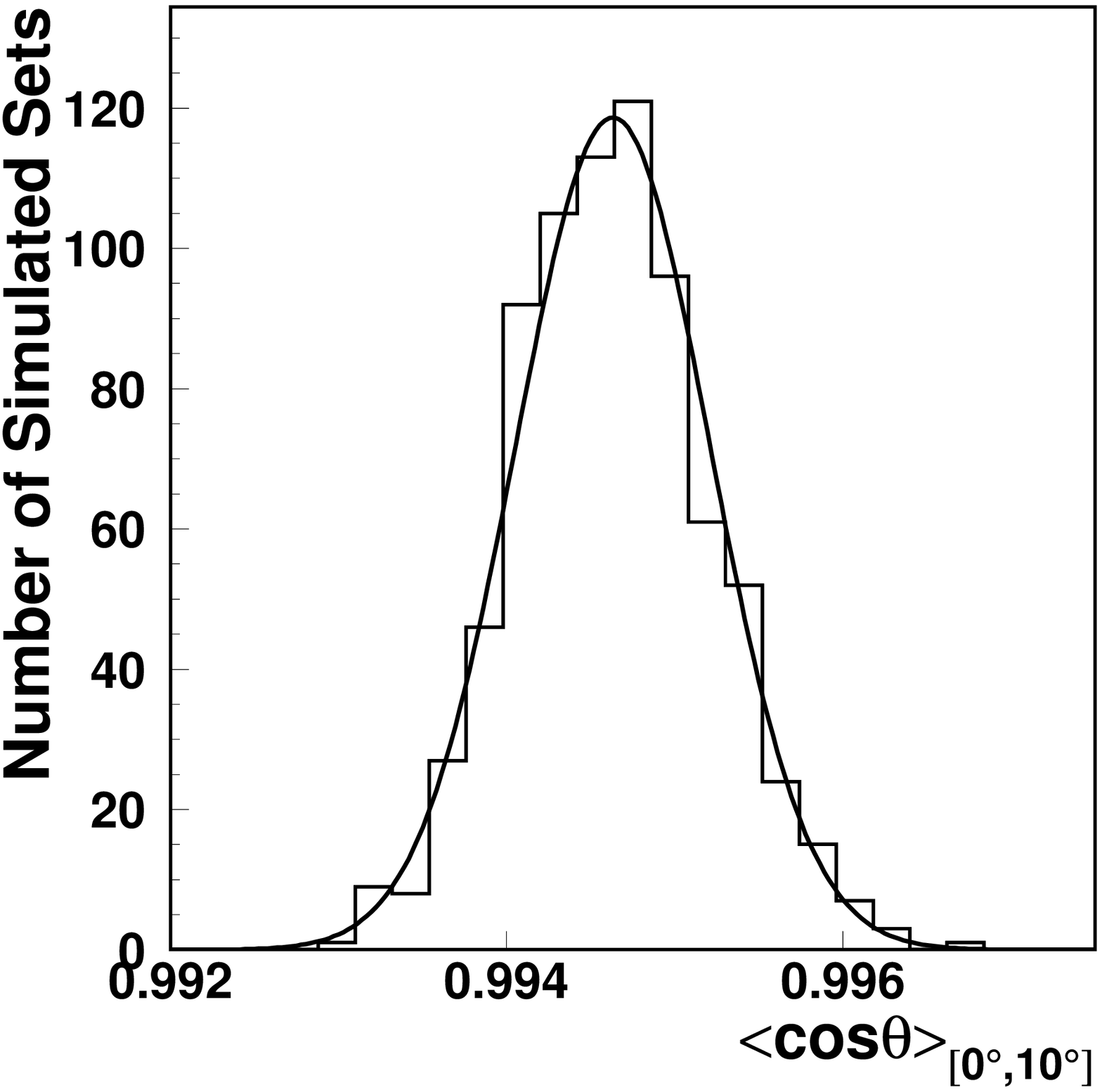}\\
\end{tabular}
\end{center}
\caption{Distributions of normalized bin densities of $\cos\theta_{\rm min}$ 
and $<\!\!\cos\theta\!\!>_{[0^\circ,10^\circ]}$ values for a large number
of simulated sets with the same level of clustering as in 
figure~\ref{fig:autoex}---(a) Distribution of observed
normalized bin densities of $\cos\theta_{\rm min}$, 
note that it is {\it not} Gaussian ($\chi^2/dof=5.44$); (b) : 
$<\!\!\cos\theta\!\!>_{[0^\circ,10^\circ]}$ distribution ($\chi^2/dof=1.09$).}
\label{fig:gauss}
\end{figure}

Figure~\ref{fig:autoex}a 
shows an example of the autocorrelation function for 
a highly clustered set of simulated data.  The sharper the peak at 
$\cos\theta_{\rm min}$ is, the
more highly autocorrelated the data set is.  There are many ways that one 
could quantify the degree of autocorrelation that a set possesses.  The 
most obvious way is to look at the value of the bin which contains
$\cos\theta_{\rm min}$.  However, this method has some pitfalls.  First of all,
the value of the last bin is dependent upon the chosen bin width.  Also, 
the value of the last bin is not stable unless the angular resolution is
sampled at a level that is computationally unfeasible. Finally, 
the value of the last bin over a large number
of similarly autocorrelated sets does {\it not} produce a Gaussian 
distribution (see figure~\ref{fig:gauss}a),  
thus complicating the interpretation of the results of an analysis employing 
$\cos\theta_{\rm min}$ as an observable.

A more well-behaved measure of the 
autocorrelation of a specific set of data is the value of
$<\!\!\cos\theta\!\!>$ for $\theta\leq10^\circ$.    
This value is also a measure of the 
sharpness of the autocorrelation peak at $\cos\theta=1$.  However, this 
method of quantification  does not
depend on bin width and it does produce Gaussian distributions 
when it is applied to large numbers of
sets with similar degrees of autocorrelation as is demonstrated in 
figure~\ref{fig:gauss}b.
An additional advantage to this method is that by considering the continuous
autocorrelation function over a specified interval, both the peak at 
the smallest values of 
$\theta$ and the corresponding statistical deficit in the autocorrelation
function at slightly higher values
of $\theta$ are taken into account.  Thus we simultaneously measure both the
positive and negative aspects of the autocorrelation signal.
The interval of 
$[0^\circ,10^\circ]$ was chosen because in simulations it was found to optimize
the autocorrelation signal for clusters resulting from point sources spread
isotropically across the sky.

\begin{figure}[t,b]
\begin{center}
\begin{tabular}{c@{\hspace{0.0cm}}c}
(a)\includegraphics[width=6.15cm]{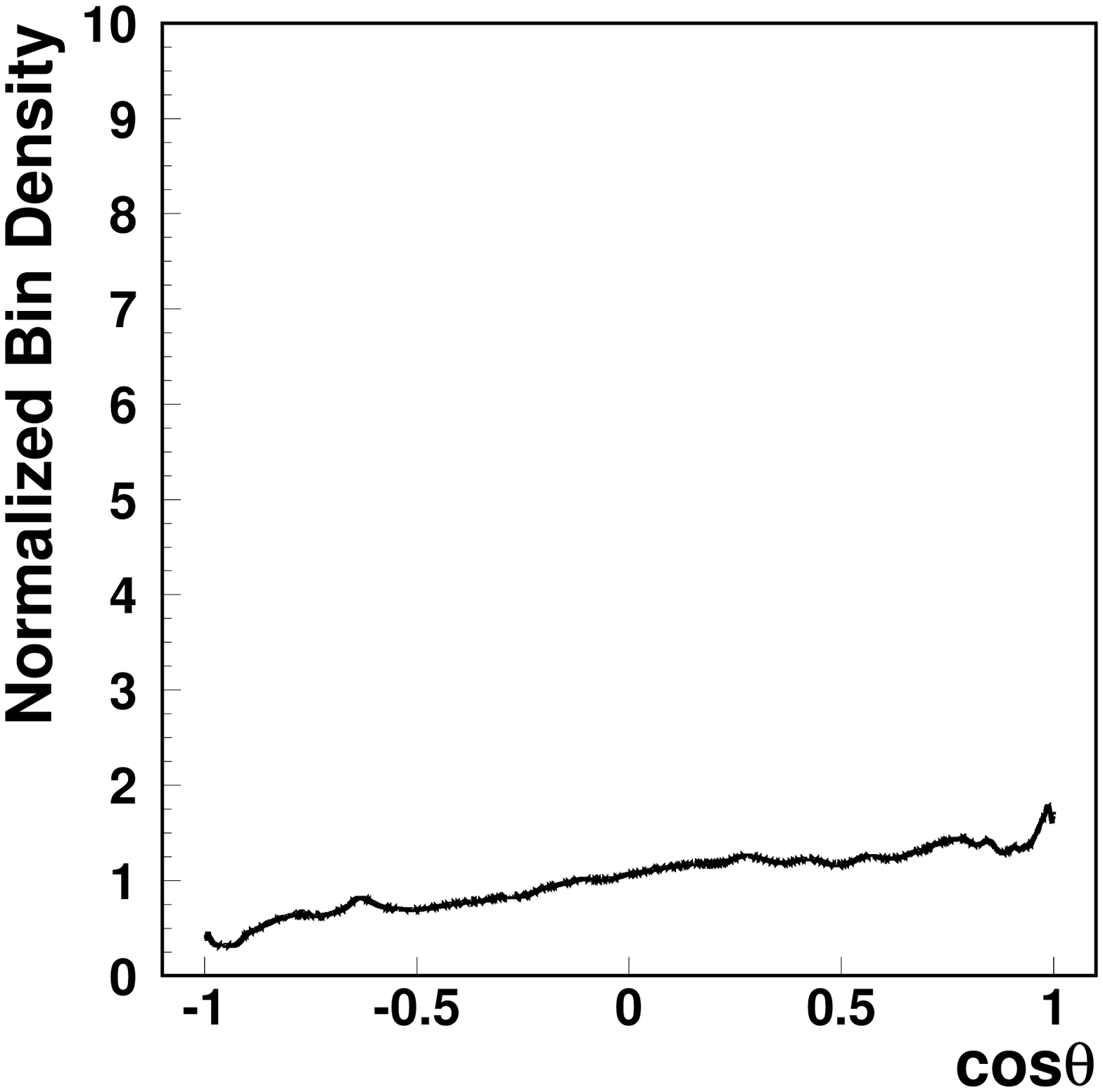}&
(b)\includegraphics[width=6.15cm]{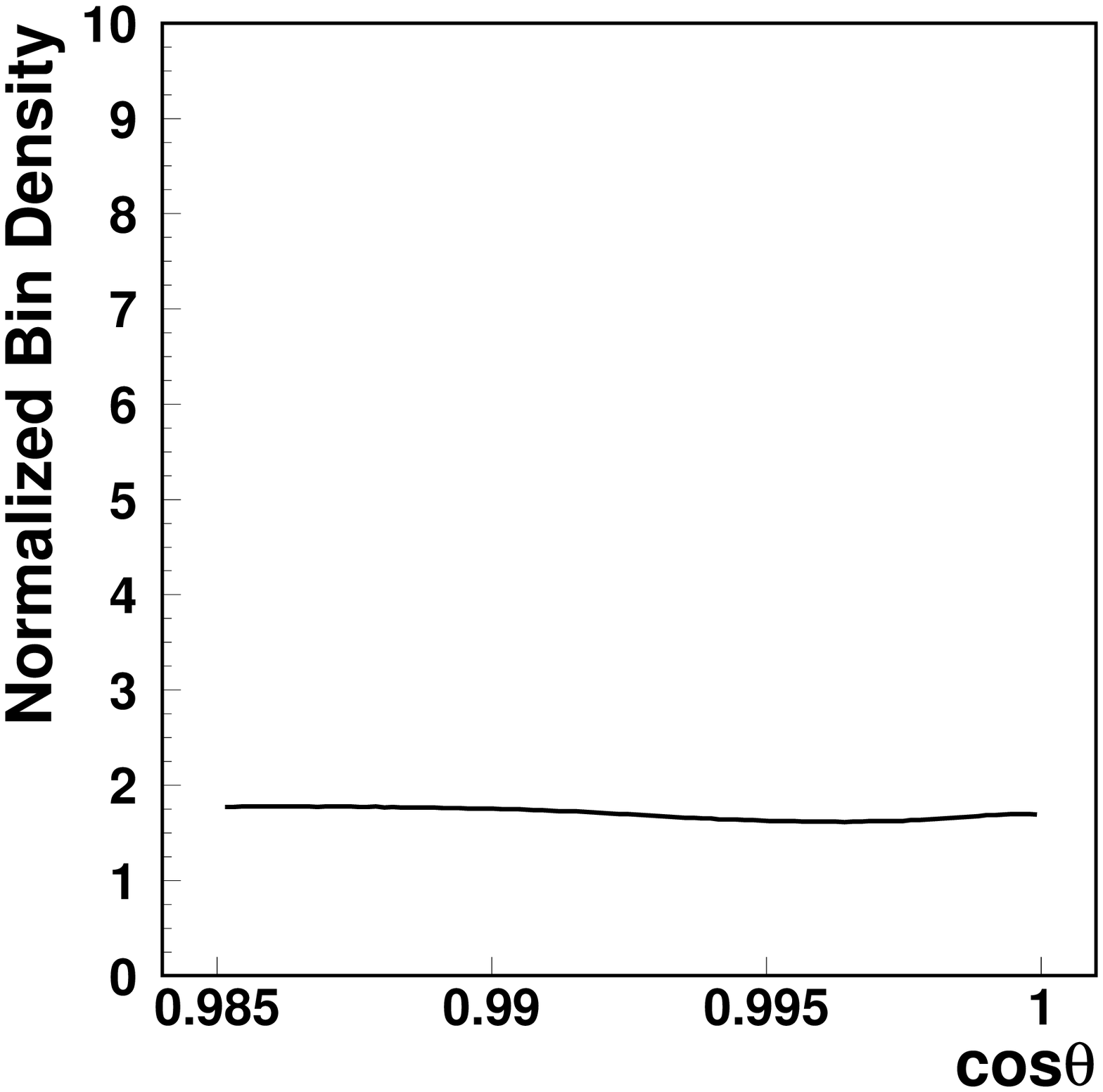}\\
\end{tabular}
\end{center}
\caption{The autocorrelation for the HiRes-I events above $10^{19.5}$~eV---(a)
the full autocorrelation function for $\theta=[0^\circ,180^\circ]$;
(b) the critical region of the autocorrelation function: 
$<\!\!\cos\theta\!\!>_{[0^\circ,10^\circ]}=0.99234$.}
\label{fig:hires_au}
\end{figure}

Using the description of the HiRes-I monocular angular 
resolution above, we then calculate the autocorrelation function via the 
method described above.  In figure~\ref{fig:hires_au}, 
we show the result of this calculation.  For this sample, we obtain 
$<\!\!\cos\theta\!\!>_{[0^\circ,10^\circ]}=0.99234$.  

\begin{figure}[t,b]
\begin{center}
\begin{tabular}{c@{\hspace{0.0cm}}c}
(a)\includegraphics[width=6.15cm]{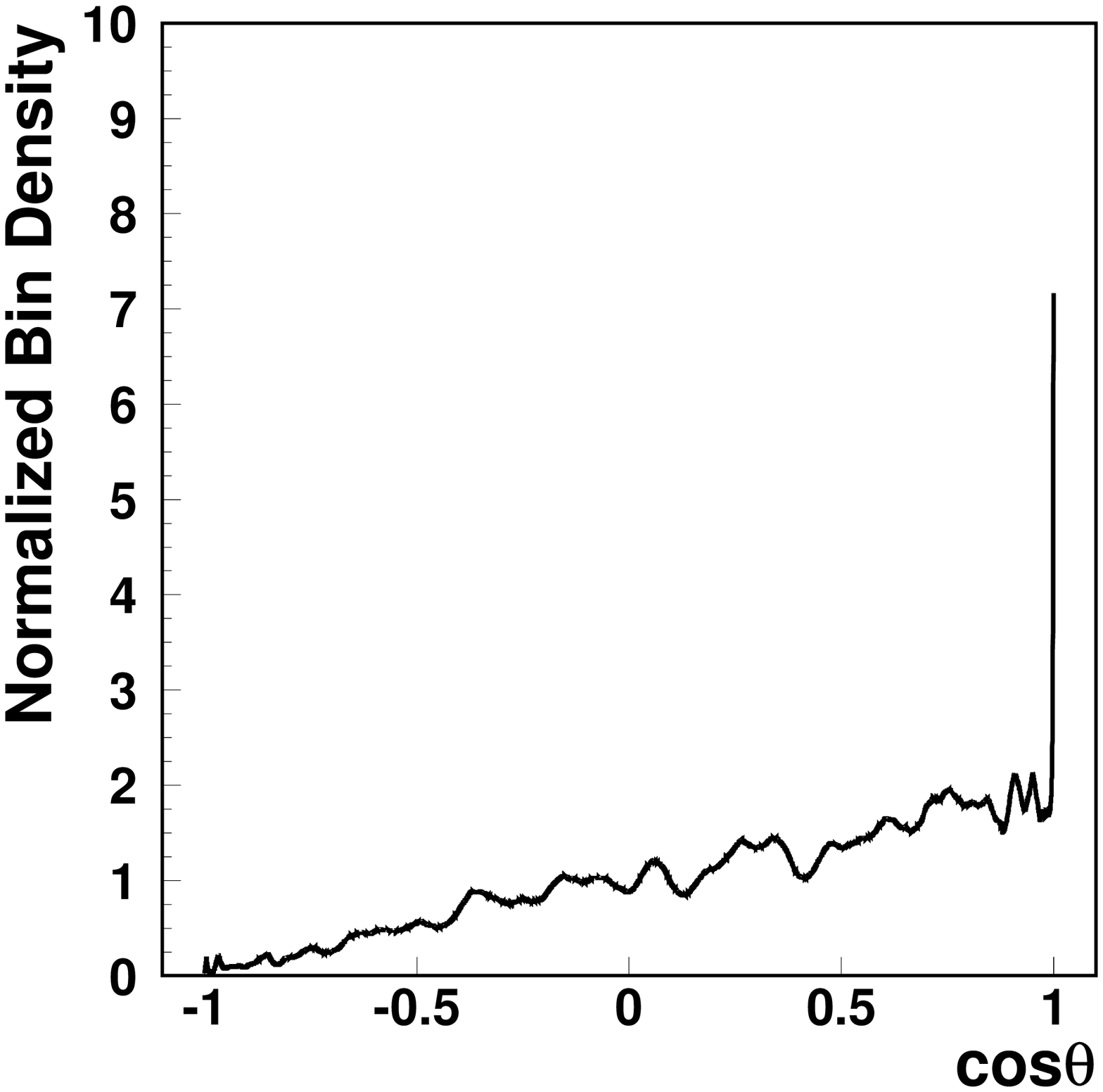}&
(b)\includegraphics[width=6.15cm]{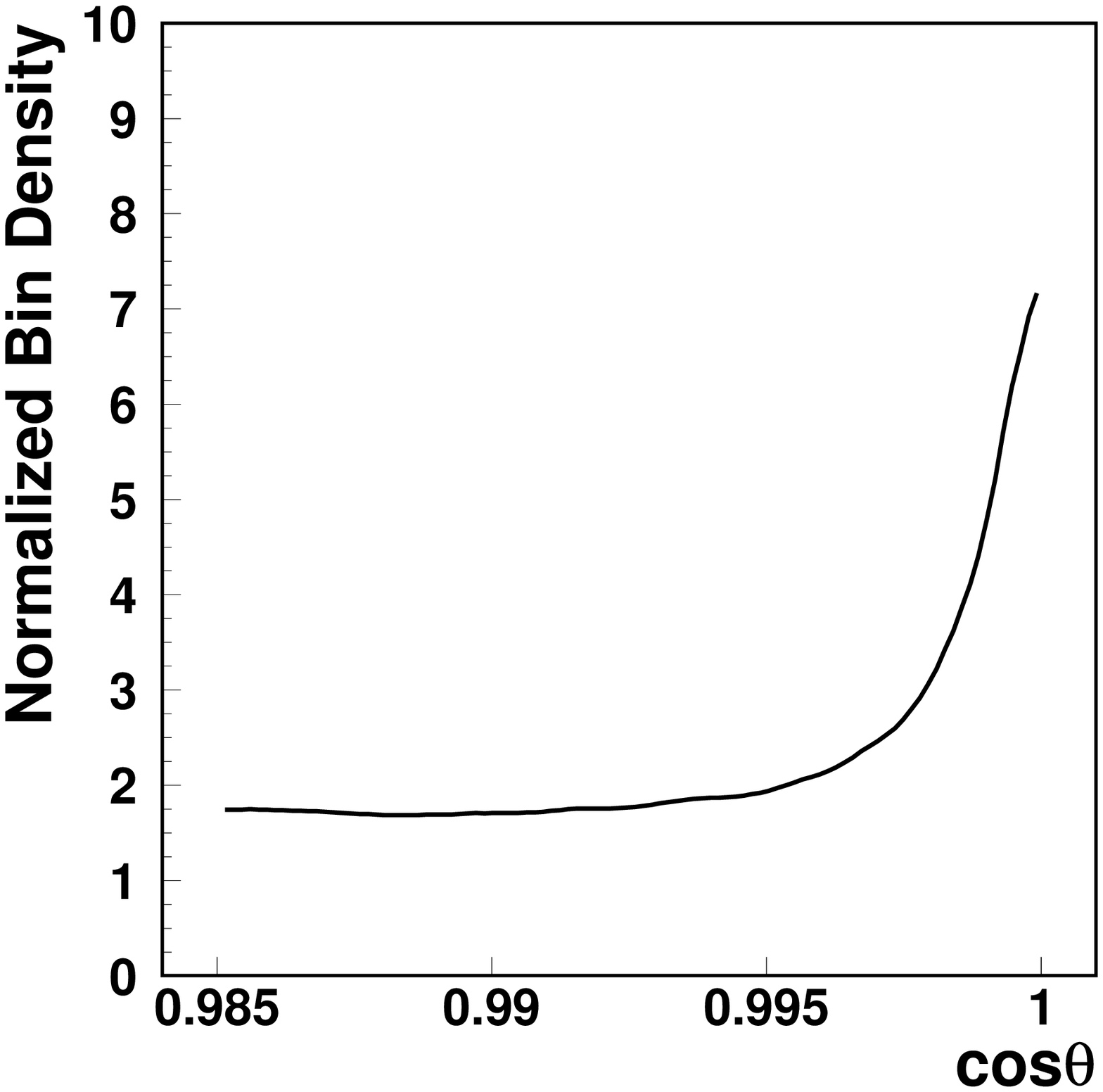}\\
\end{tabular}
\end{center}
\caption{The autocorrelation for the published AGASA events---(a) the full 
autocorrelation function for $\theta=[0^\circ,180^\circ]$; (b) the critical
region of the autocorrelation function: 
$<\!\!\cos\theta\!\!>_{[0^\circ,10^\circ]}=0.99352$.}
\label{fig:agasa_au}
\end{figure}

We also calculate the autocorrelation function for the published AGASA events.
We show the result in figure~\ref{fig:agasa_au}. 
For this sample, we obtain 
$<\!\!\cos\theta\!\!>_{[0^\circ,10^\circ]}=0.99352$.

\section{Quantifying the Relative Sensitivity of HiRes-I and AGASA to 
Autocorrelation}

\begin{figure}[t,b]
\begin{center}
\begin{tabular}{c}
\includegraphics[width=13.5cm]{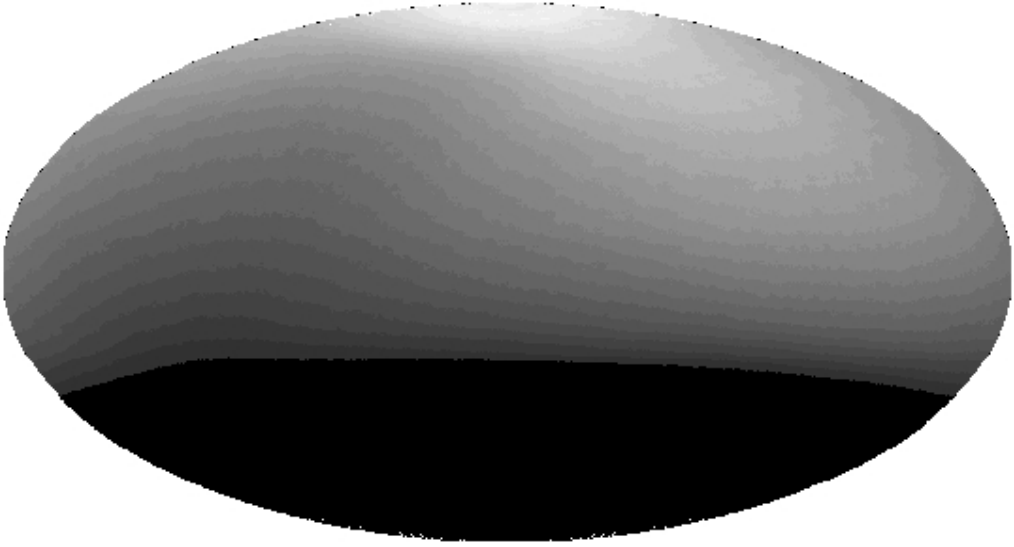}\\
\end{tabular}
\end{center}
\caption{Hires-1 estimated relative exposure, $\rho_{\rm H}(\delta,\alpha)$, 
for events above $10^{19.5}$~eV in 
equatorial coordinates (right ascension right to left).  The lightest region
corresponds to a normalized event density of 2.5.  The observable sky extends
from $\delta=-30^\circ$ to $\delta=90^\circ$.}
\label{fig:hires_ex}
\end{figure}

In order to quantify the relative sensitivity of the AGASA and HiRes-I 
data sets,
we must first understand the exposures of both detectors.
For HiRes-I, we assemble a library of approximately $8\times10^4$ 
simulated events with energies above $10^{19.5}$~eV.  We then pair each 
event with times during which the detector was 
operating.  A mirror-by-mirror 
correction is applied where simulated events
are rejected if the mirror(s) that would have observed the event in question 
was not operating at the time that event would have 
occurred.  Once $10^7$ pairings of 
simulated events and times are assembled, 
a surface plot is created of the event density on a bin by bin basis.
The value of each bin is then normalized so that the mean value of all the 
bins in the {\it observable} sky $\delta=[-30^\circ,90^\circ]$ is 1.  
The resulting surface plot is shown
in a Hammer-Aitoff projection in figure~\ref{fig:hires_ex}.
We have previously shown that this method produced zenith angle,
azimuthal angle, and sidereal time distributions that were consistent with
that observed in the actual data \cite{dipole}.  
The highest exposure areas have a normalized relative exposure:
$\rho_{\rm H}(\delta,\alpha)=\sim2.5$.

\begin{figure}[t,b]
\begin{center}
\begin{tabular}{c}
\includegraphics[width=13.5cm]{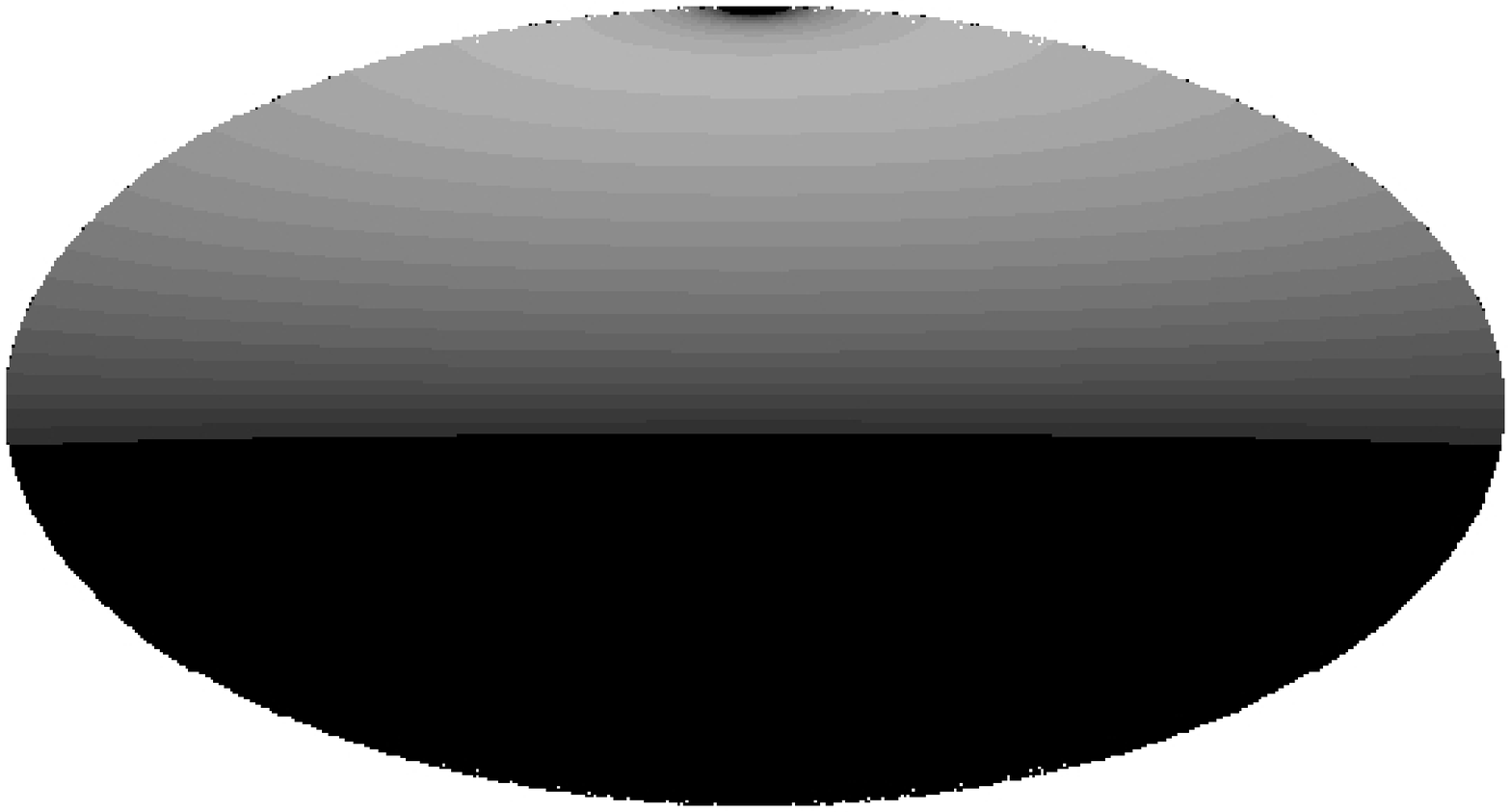}\\
\end{tabular}
\end{center}
\caption{AGASA estimated relative exposure, $\rho_{\rm A}(\delta)$, 
for events above $10^{19.5}$~eV in 
equatorial coordinates (right ascension right to left).  The lightest region
corresponds to a normalized event density of $\sim1.6$.  
The observable sky extends from $\delta=-8^\circ$ to $\delta=87.5^\circ$.}
\label{fig:agasa_ex}
\end{figure}

\begin{figure}[t,b]
\begin{center}
\begin{tabular}{c@{\hspace{0.0cm}}c}
(a)\includegraphics[width=6.15cm]{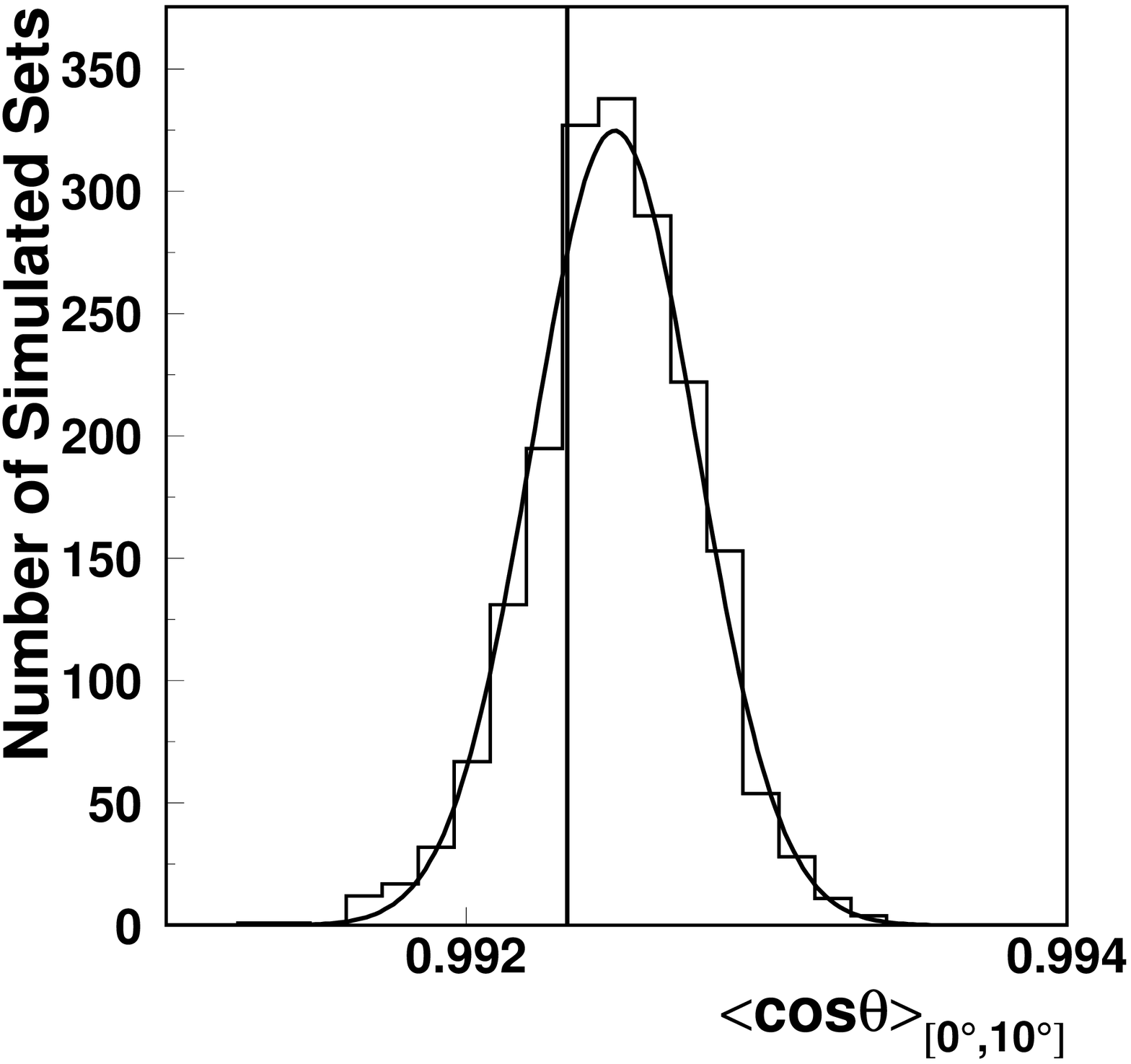}&
(b)\includegraphics[width=6.15cm]{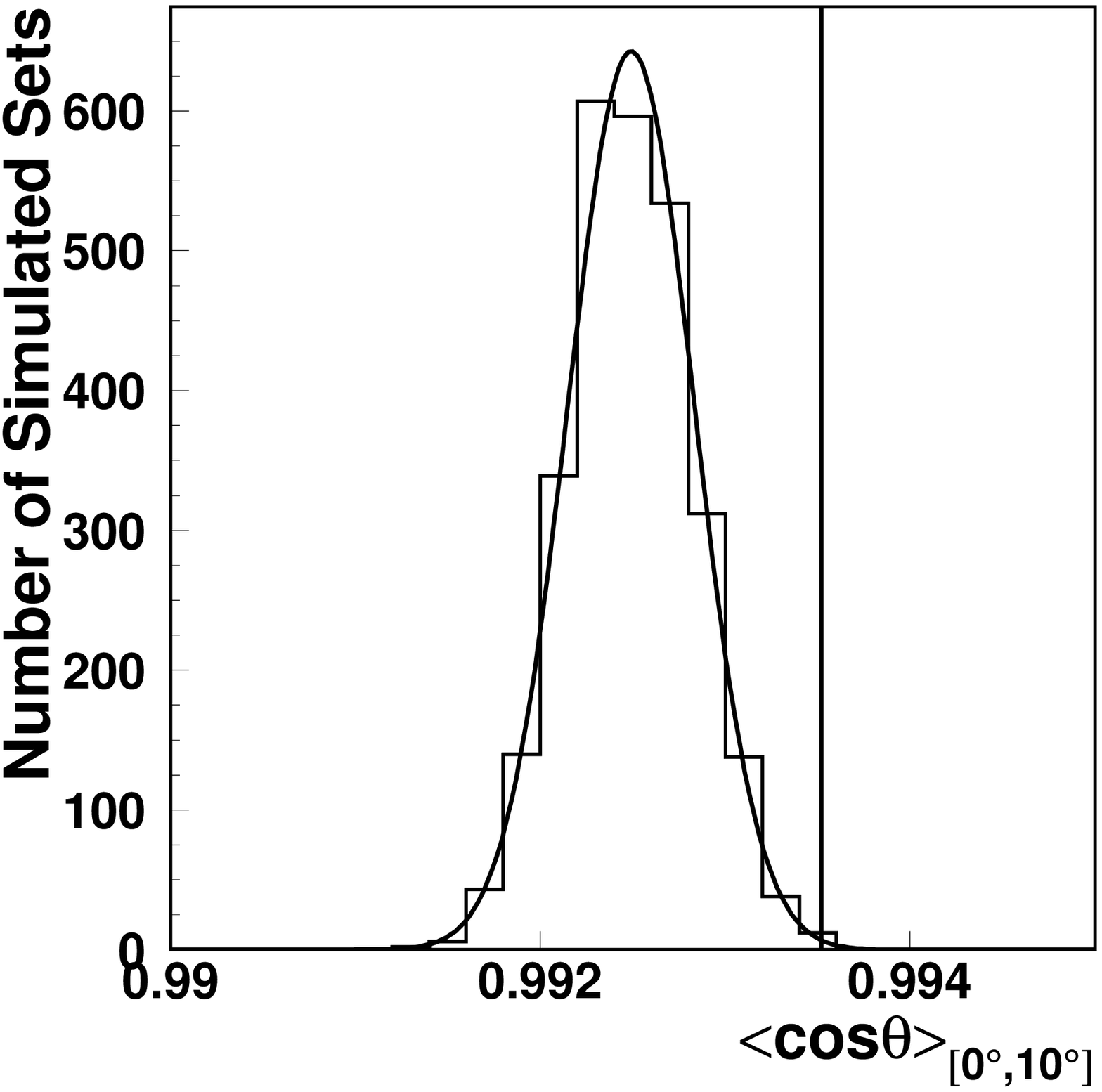}\\
\end{tabular}
\end{center}
\caption{Distribution of $<\!\!\cos\theta\!\!>_{[0^\circ,10^\circ]}$ values
for simulated isotropic data sets---(a) HiRes-I; (b) AGASA.  In each figure,
the vertical line represents the the value of 
$<\!\!\cos\theta\!\!>_{[0^\circ,10^\circ]}$ for the observed data.}
\label{fig:iso}
\end{figure}

For the AGASA detector, we refer to the distribution of event declinations
presented in Uchiori {\it et al.}  \cite{Uchihori:1999gu}.
By following the lead of Evans {\it et al.} \cite{Evans:2001rv}, 
we fit a normalized polynomial to this distribution:
\begin{eqnarray}
N(\delta) & = & 0.323616+0.0361515\delta-5.04019\times10^{-4}\delta^2+ \nonumber \\
          &   & 5.539141\times10^{-7}\delta^3;
\end{eqnarray}
where $N(\delta)$ holds for $\delta=[-8^\circ,87.5^\circ]$the maximum value 
of $N(\delta)$ is 1.  We also know that:
\begin{equation}
A_\circ\int_{-8^\circ}^{87.5^\circ}N(\delta) \, d\delta = 
\int_{-8^\circ}^{87.5^\circ}\rho_{\rm A}(\delta)\cos\delta \, d\delta,
\end{equation}
where $A_\circ$ is a numerically determined normalization constant.  We 
then derive:
\begin{equation}
\rho_{\rm A}(\delta)= A_\circ N(\delta)\sec\delta; A_\circ=1.0251.
\label{eqn:rhoa}
\end{equation}
The value of each bin is once again normalized 
so that the mean value of all the 
bins in the {\it observable} sky $\delta=[-8^\circ,87.5^\circ]$ 
is 1.  The resulting surface plot is shown
in a Hammer-Aitoff projection of a equatorial coordinates in 
figure~\ref{fig:agasa_ex}.
The highest exposure areas have $\rho_{\rm A}(\alpha)=\sim1.6$. 
In figure \ref{fig:iso},
we show the distribution of $<\!\!\cos\theta\!\!>_{[0^\circ,10^\circ]}$
values for isotropic data sets with each of the two different exposure models
(HiRes-I and AGASA).  The AGASA data set manifests $\sim10^{-3}$ chance 
probability above background.  For the AGASA data, 
we also calculated the autocorrelation function
without consideration to angular resolution and employed the more conventional
$\theta_{min}$ observable.  After varying the bin width for 
$\theta_{min}$ and accounting for the trials factor, we independently concluded
that the chance probability is $\sim10^{-3}$ for the optimal bin width, 
$\theta_{min}=[0^\circ,2.5^\circ]$.  We thus conclude that factoring angular
resolution into our analysis and employing 
$<\!\!\cos\theta\!\!>_{[0^\circ,10^\circ]}$ as an observable in no way 
diminishes the sensitivity to autocorrelation in the reported AGASA data.

There are a few important differences between the exposure of the HiRes-I and 
AGASA detectors.  First of all, the exposure of the HiRes-I detector is 
more asymmetric than the exposure of the AGASA detector.  This is not only
due to seasonal variations in the HiRes detector, but also due to its ability
to constantly observe the region around $\delta=90^\circ$ due to a higher
zenith angle acceptance.  This higher zenith angle acceptance also allows
the HiRes detector to observe a greater region of the southern hemisphere.
In general, while AGASA reports 
observations for 56.9\% of the total sky, the HiRes-I detector reports 
observations for 75\% of the total sky.

To simulate clustering we use the following prescription:  
\begin{enumerate}
\item An event is chosen based upon the distribution in $\alpha$ and $\delta$
that is dictated by $\rho$.  
In the case of HiRes-I, this is simply done
by selecting a simulated event from our library and then assigning it a time
that is a known good-weather ontime for the mirror(s) that observed
that event.  In the case of the AGASA detector, this is done by selecting a 
random value for $\delta$ that conforms to the distribution in 
equation~(\ref{eqn:rhoa}) 
and then assigning it a random value in $\alpha$ between
0h and 24h and sampling a value for the energy from the energies of the 
reported events.
\item This event does not represent the source location itself, but
is assumed to have arrived from the source location with some error.  We
construct a "true" source location by sampling the error space of this
event.
\item For each additional event assigned to that source, a simulated event is
selected with a ``true'' arrival direction that is the same as that of the 
initial event.
\end{enumerate}

\begin{figure}[t,b]
\begin{center}
\begin{tabular}{c@{\hspace{0.0cm}}c}
(a)\includegraphics[width=6.15cm]{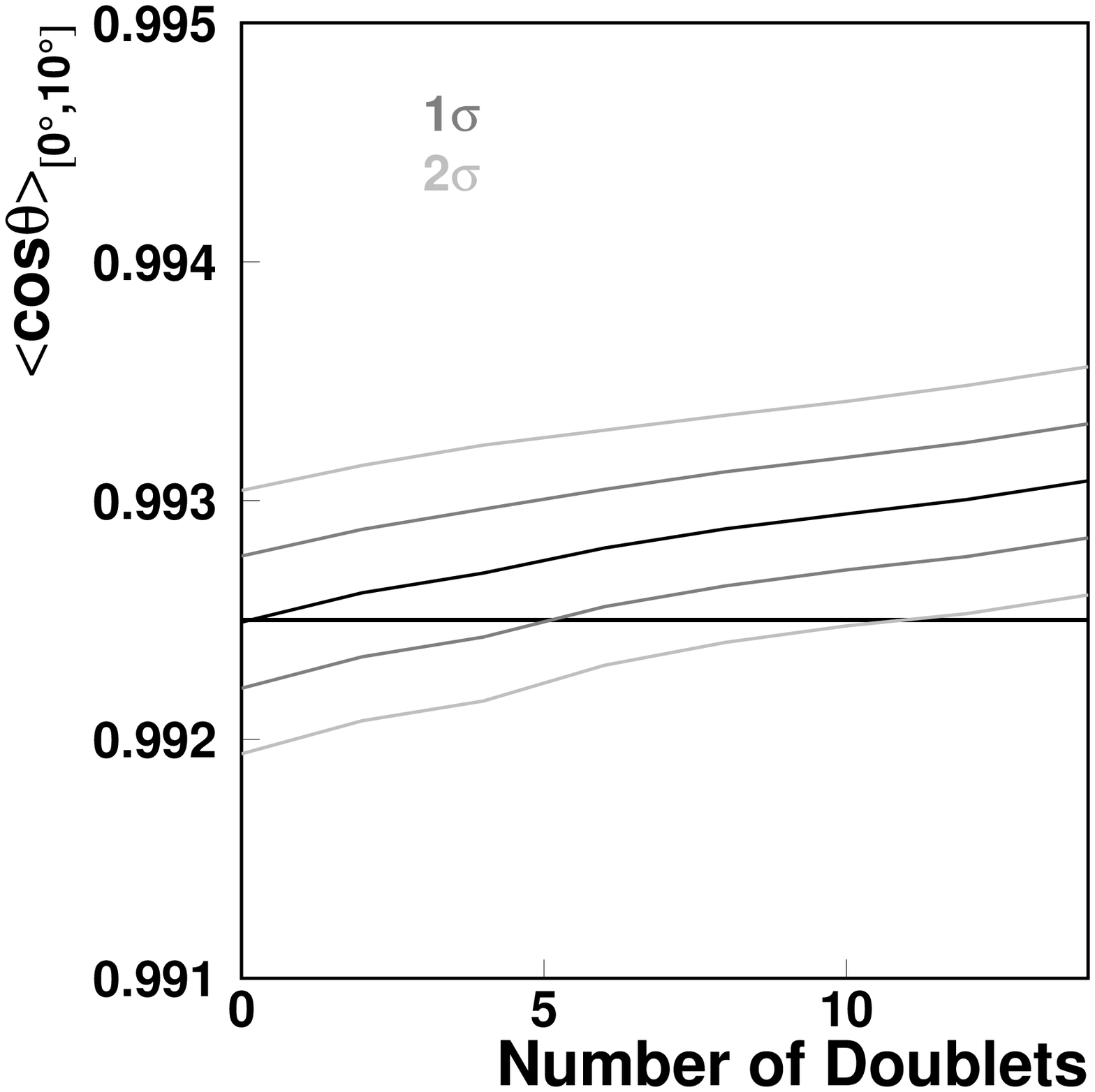}&
(b)\includegraphics[width=6.15cm]{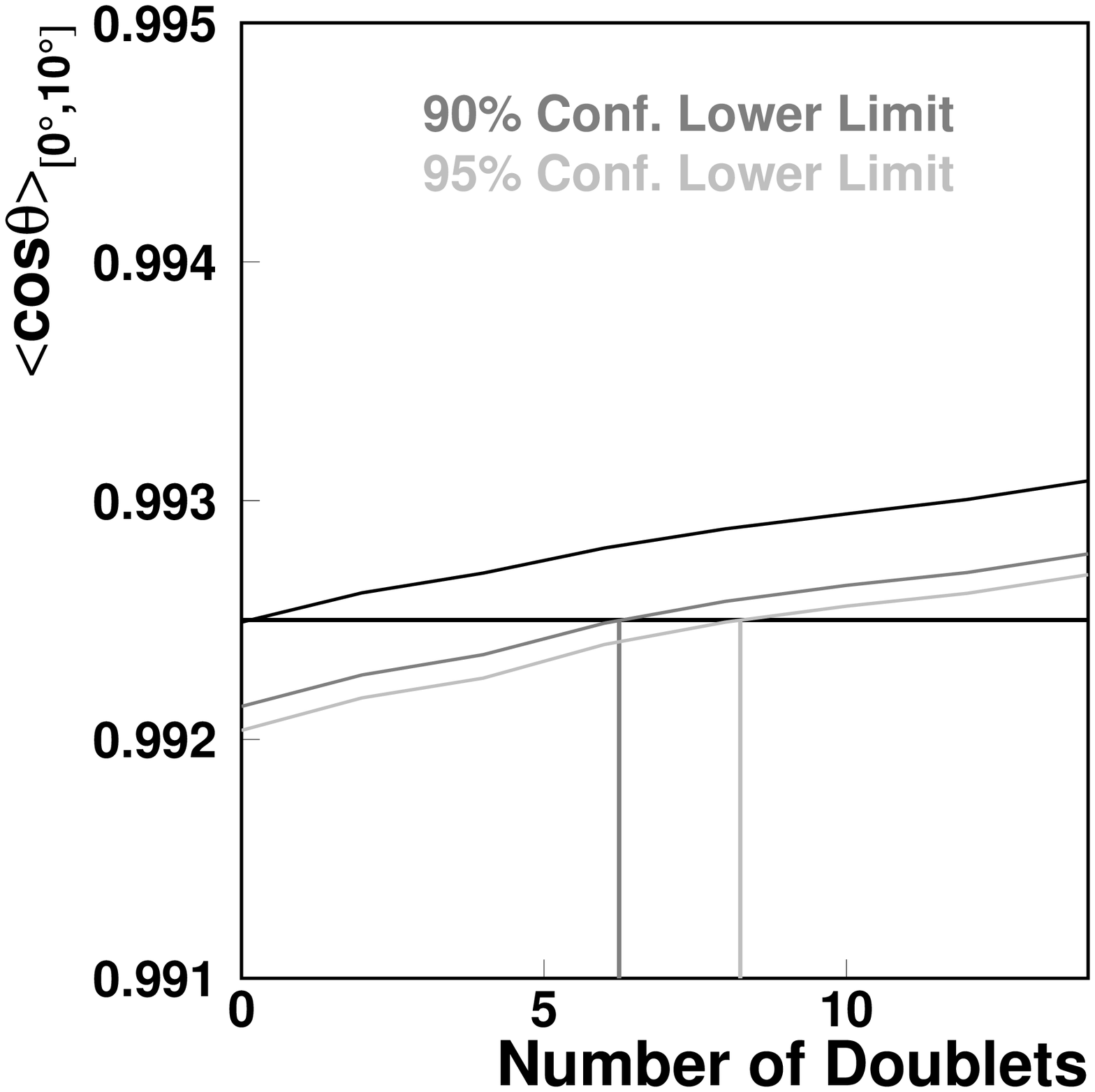}\\
(c)\includegraphics[width=6.15cm]{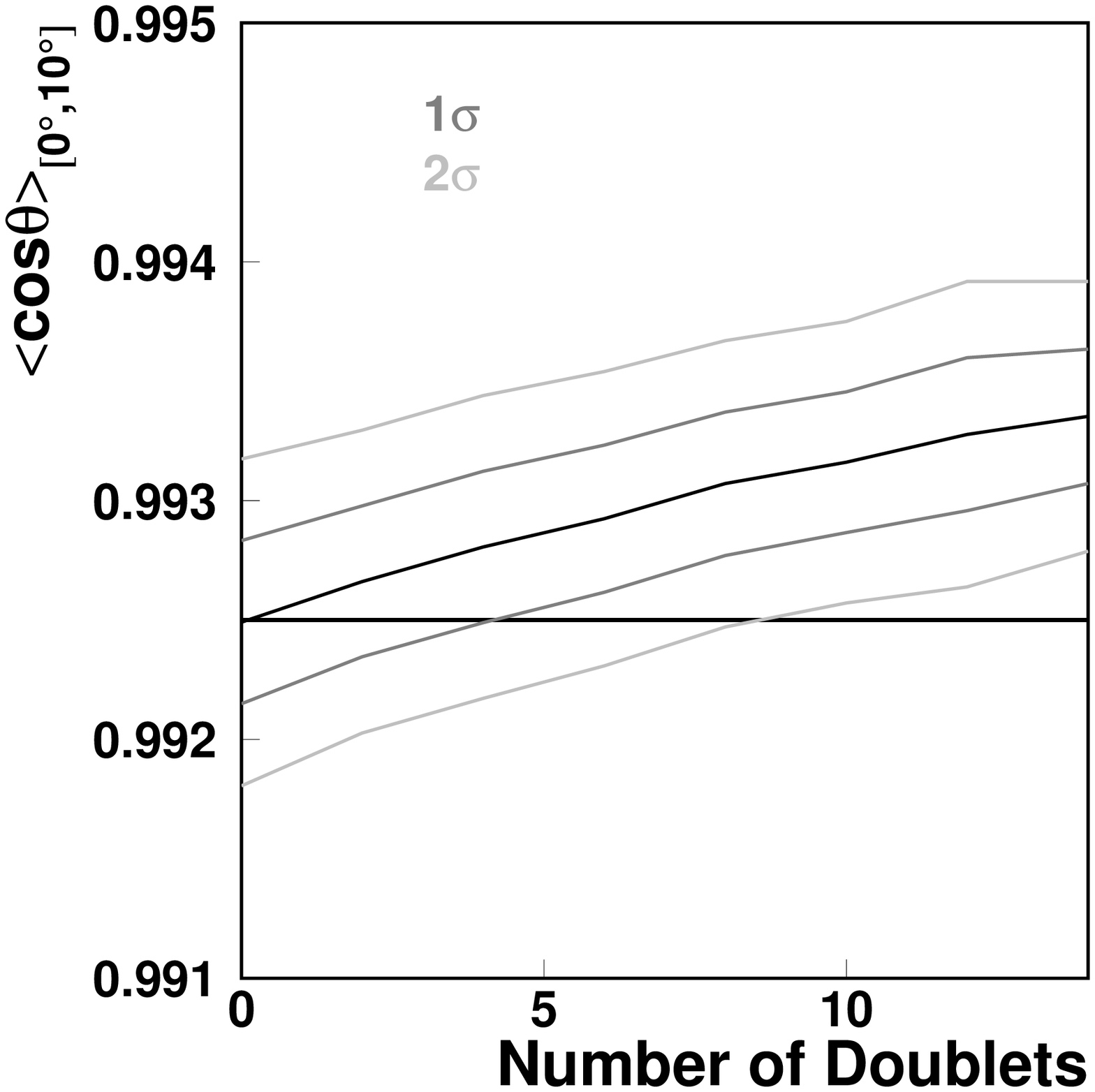}&
(d)\includegraphics[width=6.15cm]{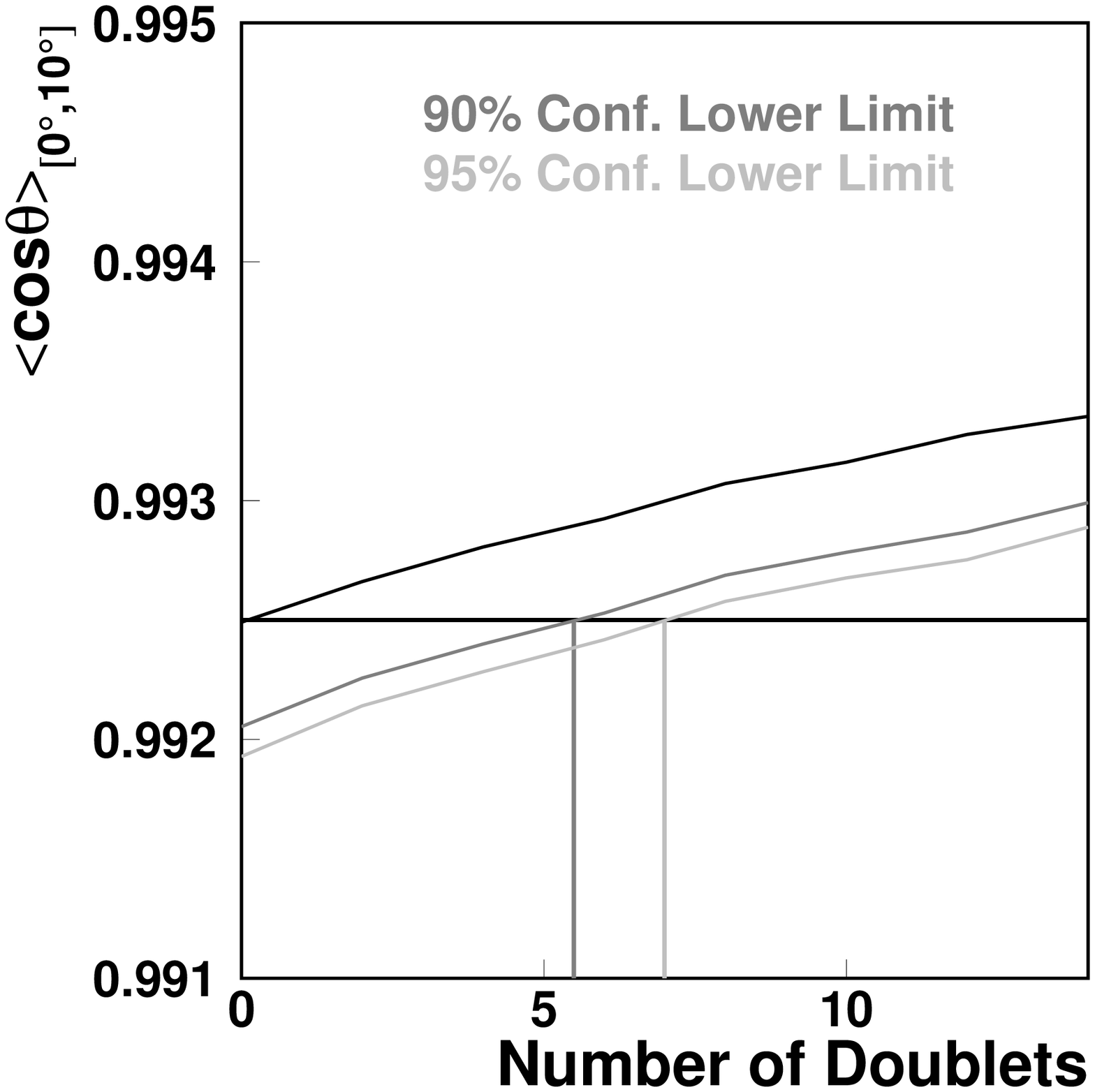}\\
\end{tabular}
\end{center}
\caption{Relative sensitivity of HiRes-I and AGASA to doublets---(a) 
Simulations with the HiRes-I detector and 52 events; 
(b) 90\% confidence above background: 6.25 doublets, 95\% confidence above 
background: 8.25 doublets.
(c) simulations with the AGASA detector and 59 events; 
(d) 90\% confidence above background: 5.5 doublets, 
95\% confidence above background: 7.0 doublets.
In each figure, the horizontal line indicates the expected value of 
$<\!\!\cos\theta\!\!>_{[0^\circ,10^\circ]}$ for an isotropic background}
\label{fig:doublets}
\end{figure}

To study the relative sensitivity of AGASA and HiRes-I, we 
measure the value of $<\!\!\cos\theta\!\!>_{[0^\circ,10^\circ]}$ for multiple
simulated sets with a variable number of doublets inserted.  We then 
construct an interpolation of the mean value and standard deviation of
$<\!\!\cos\theta\!\!>_{[0^\circ,10^\circ]}$ from a given number of observed
doublets for each experiment.   This will allow us to state the number of
doublets required for each experiment in order for the 90\% confidence 
limit of 
$<\!\!\cos\theta\!\!>_{[0^\circ,10^\circ]}$ to be above the background value
of 0.99250.  Figure~\ref{fig:doublets} 
shows the result
of these simulations.  In general, for the HiRes-I data set, the 90\% 
confidence lower limit corresponds to the mean expected background signal
with the inclusion of 6.25 doublets.  For AGASA,the 90\% 
confidence lower limit corresponds to the mean expected background signal
with the inclusion of 5.5 doublets. 
This demonstrates that while AGASA has a slightly better ability to perceive 
autocorrelation, the sensitivity of the two experiments is comparable.

\begin{figure}[t,b]
\begin{center}
\begin{tabular}{c@{\hspace{0.0cm}}c}
(a)\includegraphics[width=6.15cm]{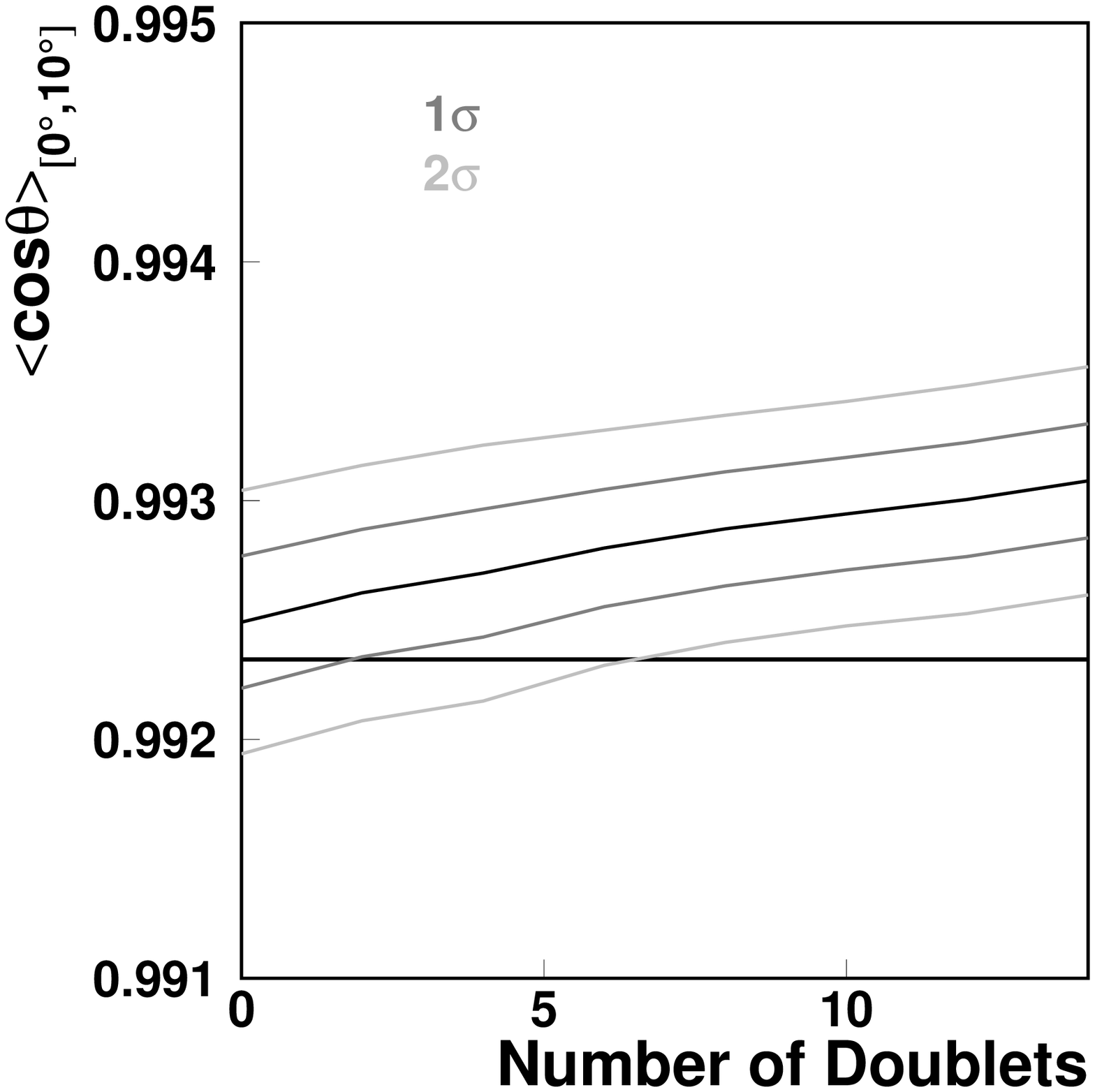}&
(b)\includegraphics[width=6.15cm]{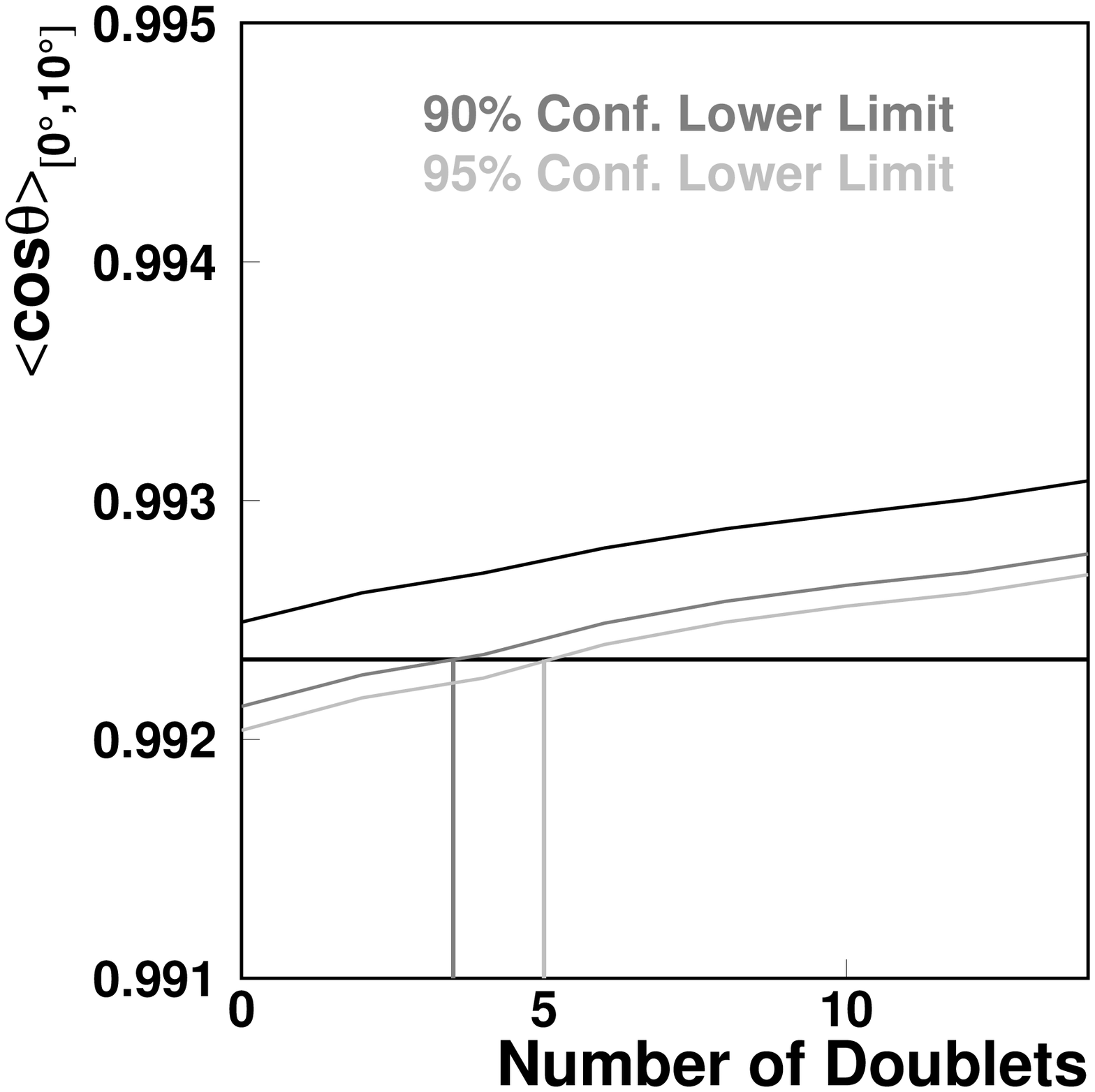}\\
\end{tabular}
\end{center}
\caption{Sensitivity of the HiRes-I monocular observations to doublets---(a) 
Simulations with the HiRes-I detector and 52 events; 
(b) 90\% confidence above observed signal: 3.5 doublets, 95\% confidence above 
observed signal: 5 doublets.  In each plot, the horizontal line represents
the value of $<\!\!\cos\theta\!\!>_{[0^\circ,10^\circ]}$ for the observed 
HiRes-I data}
\label{fig:doublets2}
\end{figure}

We now apply the actual Hires-I $<\!\!\cos\theta\!\!>_{[0^\circ,10^\circ]}$ 
to the sensitivity curve shown in figure~\ref{fig:doublets}.
In figure~\ref{fig:doublets2} 
we can see the result
of these simulations.  The observed HiRes-I signal corresponds
to the  90\% confidence upper limit with the inclusion of only 3.5 doublets
beyond random background coincidence.  

If we repeat this analysis with first, a 7.5\% reduction in the estimated 
angular resolution values and second, a 7.5\% increase in the estimated 
angular resolution values, we obtain a range for the 90\% confidence upper 
limit of $[2.75, 4.0]$ doublets and a range for the 95\% confidence upper 
limit of $[4.5, 5.5]$ doublets.  

A final area of concern is the systematic uncertainty in the 
determination of atmospheric clarity.  Because hourly atmospheric observations
are not available for the entire HiRes-I monocular data set, we
have relied upon the use of an average atmospheric profile for the 
reconstruction of our data \cite{Wiencke:atmos}.  
While different atmospheric conditions have negligible impact on the 
determination of the arrival direction for events with measured energies this
high, differing conditions can have an impact on energy estimation and thus
the number of events that are included in our data set.  Over the $1\sigma$ 
error space for our estimation of atmospheric conditions, the total number
of events in our data set fluctuates on the interval $[41,65]$.  The value of
the observable,  $<\!\!\cos\theta\!\!>_{[0^\circ,10^\circ]}$, has a fluctuation
on the interval $[0.99226,0.99249]$ owing to addition and subtraction of 
events from the data set.  Note that in neither case does the value
of $<\!\!\cos\theta\!\!>_{[0^\circ,10^\circ]}$ exceed the mean value (0.99250)
expected for a background set.
\section{Conclusion}

We conclude that the HiRes-I monocular detector sees no 
evidence of clustering in its highest energy events.  Furthermore, the HiRes-I
monocular data has an intrinsic sensitivity to global autocorrelation such that
we can claim at the 90\% confidence level that there can be no more than 3.5 
doublets above that which would be expected by background coincidence
in the HiRes-I monocular data set above $10^{19.5}$~eV.  From this result,
we can then derive,
with a 90\% confidence level, that no more than 13\% of the observed HiRes-I
events could be sharing common arrival directions. This data set is
comparable to the sensitivity of the reported AGASA data set if one
assumes that there is indeed a 30\% energy scale difference between the two
experiments.  It should be 
emphasized that this conclusion pertains only to point sources of the sort
claimed by the AGASA collaboration.  Furthermore, because a
measure of autocorrelation makes no assumption of the underlying astrophysical
mechanism that results in clustering phenomena, we cannot claim that the HiRes 
monocular analysis and the AGASA analysis are inconsistent beyond a specified 
confidence level. 

\section{Acknowledgments}
This work is supported by US NSF grants PHY 9322298, PHY 9321949, 
PHY 9974537, PHY 0071069, PHY 0098826, PHY 0140688, PHY 0245428, PHY 0307098
by the DOE grant FG03-92ER40732,
and by the Australian Research Council. We gratefully
acknowledge the contributions from the technical staffs of our home
institutions. We gratefully acknowledge the contributions from the University
of Utah Center for High Performance Computing. The cooperation of 
Colonels E. Fisher and G. Harter, the US Army and the Dugway Proving Ground 
staff is appreciated.

\end{document}